\documentclass[%
 reprint,
 amsmath,amssymb,
 aps,
]{revtex4-2}

\usepackage{graphicx}
\usepackage{dcolumn}
\usepackage{bm}
\usepackage{ulem}
\usepackage{xcolor} 

\begin{document}

\preprint{APS/123-QED}

\title{Motivation and needs of informal physics practitioners}

\author{Shams El-Adawy}
\affiliation{Department of Physics, Kansas State University, Manhattan, KS 66506}

\author{Alexandra C. Lau}
\affiliation{American Physical Society, 1 Physics Ellipse, College Park, MD 20740 }

\author{Eleanor C. Sayre}
\affiliation{Department of Physics, Kansas State University, Manhattan, KS 66506}
\affiliation{Center for Advancing Scholarship to Transform Learning, Rochester Institute of Technology, Rochester, NY 14623}

\author{Claudia Fracchiolla}
\affiliation{American Physical Society, 1 Physics Ellipse, College Park, MD 20740}

\date{\today}

\begin{abstract}
Physicists engage with the public to varying degrees at different stages of their careers. However, their public engagement covers many activities, events, and audiences, making their motivations and professional development needs not well understood. As part of ongoing efforts to build and support community in the informal physics space, we conducted interviews with physicists with a range of different experiences in public engagement. We use personas methodology and self-determination theory to articulate their public engagement motivation, challenges, and needs. We present our set of three personas: the physicist who engages in informal physics for self-reflection, the physicist who wants to spark interest and understanding in physics, and the physicist who wants to provide diverse role models to younger students and inspire them to pursue a STEM career. Needs covered a range of resources including science communication training, community building among informal physics practitioners, and mechanisms to recognize, elevate and value informal physics. By bringing user-centered design methodology to a new topical area of physics education research, we expand our understanding of motivations and needs of practitioners in physics public engagement. Therefore, departments, organizations and institutions could draw upon the personas developed to consider the ways to better support physicists in their respective environment.
\end{abstract}

\maketitle


\section{Introduction}
Informal physics education refers to activities and events centered on engagement with
physics outside the formal classroom. Public engagement has been defined as encompassing “the myriad of ways in which the activity and benefits of higher education and research can be shared with
the public. Engagement is by definition a two-way process, involving interaction and
listening, with the goal of generating mutual benefit”\cite{national_coordinating_center_for_public_engagment_what_2014}.We refer to informal physics and public engagement interchangeably as informal physics activities play an important role in the public’s general understanding of physics and
science. 

Many types of activities, platforms and programs fall under informal physics education
such as after-school programs, public talks, demonstration presentations, open houses,
science festivals, planetariums, social media, websites, popularized books, movies and
games \cite{hein2009learning}. While many of these activities can be specific to physics and astronomy, some of them include a broader sense of education across science fields or all of STEM.  Despite the wide variety of possible activities, a common characteristic they
share is that participation is voluntary and activities are meant to provide participants
the freedom to explore and be curious about how the world works.

Research in informal physics, often referred to as IPER,  has focused on physics identity development, development of informal education programs, skill development for facilitators, impact of engagement in informal physics on audiences and the landscape of practices undertaken in this space \cite{tacsar2023international}. 
Research shows that participation in informal physics programs significantly enhances facilitators' communication skills, teamwork capacity and confidence   \cite{wulf2013promoting, hinko2013impacting, rethman2021impact}. Moreover, participation in these programs has the added benefit of increasing sense of belonging to the field of physics for both facilitators and audience. In particular, for individuals from underrepresented populations, engagement with physics in these informal spaces allows them to develop their physics identity as they bring their whole selves to these spaces \cite{fracchiolla2016university, wulff2018engaging, bell2019informal, fracchiolla2020community,rethman2021impact, prefontaine2021informal}. In turn, informal physics increases the interest and relevance of physics and science as a potential career path \cite{allen2017afterschool}. 
 
Furthermore, informal education programs provide opportunities for significant numbers of individuals in various geographic locations and diverse demographics to hear and engage with physics and physicists \cite{izadi2022towards}. The dimensions at play in informal physics programs are varied, rich and nuanced.  In a study about the landscape of informal physics Izadi \textit{et al.} provide an overview of all possible components of informal programs: personnel (volunteers and paid staff), resources (funding and community partners), program (goals, interactive activities and physics content), audience (geographic location and attendee demographics), assessment (educational research, tools and instruments for evaluation), and institution (role of institution administration and type of host institution)\cite{izadi2022towards}. These various dimensions show the different avenues and challenges to engage with audiences about physics. 

Efforts have also been made to survey programs to characterize some of the challenges faced in this space \cite{fracchiolla2019characterizing, izadi2019developing}. Factors such as personnel \cite{stanleycentral} and funding were among the biggest barriers to the functionality and sustainability of programs long-term \cite{fracchiolla2019characterizing, izadi2019developing}. Additionally, there is a common sense of isolation for facilitators and researchers in informal physics education who struggle to sustain and grow their efforts in informal spaces \cite{thorley2016physicists,bennett2021challenges}. Nevertheless, research remains scarce on the needs of facilitators of informal physics activities. 
Given that there is little research on what type of training and support these practitioners and researchers need in order to sustain, grow, and feel connected to a community of informal science educators, it is necessary to better understand the experiences of the physicists who facilitate these informal activities.

The central professional organization in the field of physics, the American Physical Society (APS) \cite{aps_mission_2022}, has implemented initiatives and programs to engage the physics community in public engagement \cite{aps_mission_2022}.  The APS  Public Engagement unit creates programs and professional development to support physics learning outside the traditional formal classroom. The Public Engagement unit aims to support and empower the physics community to promote access and widespread participation in physics through informal physics education activities. 
In the first half of 2022, APS Public Engagement aimed to gather members’ needs and interests around public engagement in order to inform the development of an initiative named The Joint Network for Informal Physics Education and Research (JNIPER) \cite{aps_public_engagement_joint_2022}, pronounced ``Juniper''. Gathering data on members' needs led to the development of JNIPER and development of a survey of APS members’ involvement, interests, challenges, and perceived value regarding public engagement in physics contracted with the American Institute of Physics Statistical Research Center \cite{aps_public_engagement_aps_2023}.

\subsection{Joint Network for Informal Physics Education and Research}

JNIPER brings together  physicists who facilitate informal physics learning activities, along with researchers who investigate the impact of these activities, to align and centralize the informal learning efforts of the physics community at large \cite{aps_public_engagement_joint_2022}. JNIPER aims to contribute to broad success of informal physics programs by creating a centralized network that meets three major goals: 
\begin{enumerate}
    \item Create a supportive, foundational community that connects groups of researchers and practitioners of informal physics education \cite{lau_jniper_2022};
    \item Facilitate research-practice partnerships to advance knowledge within the informal physics education research community \cite{lau_jniper_2022}; 
    \item Support adoption of research-based informal physics education best practices \cite{lau_jniper_2022}.
\end{enumerate}

The long-term vision of JNIPER is to elevate the value and recognition for public engagement in physics. The network seeks to broaden participation in physics by fostering public engagement programs that are grounded in research and cultural competency, and oriented towards equity \cite{lau_jniper_2022}. The leadership team behind the JNIPER project defines culturally-competent practices in informal physics programs as practices that are knowledgeable of and responsive to the values, desires and practices of the community they are engaged with \cite{moule2011cultural, lau_jniper_2022}. Equity-oriented practices are practices that ensure everyone has the same opportunities to engage in the environment considering privileges and power dynamics at play \cite{godec2022interested}. The mission of the JNIPER initiative is to empower and support informal physics practitioners and researchers to enact these best practices. 

However, since the pathways and engagement of physicists in informal physics education are varied, the first author, alongside the APS Public Engagement team, developed a research project to better understand physicists’ needs around public engagement by leveraging APS's membership network. We were guided by the goal to design research-based, useful programming  to support physicists' involved in public engagement. We sought to answer: \textbf{What are the motivations and professional development needs of physicists who engage in informal physics?} To answer this question, we conducted interviews with physics practitioners and researchers with a range of different experiences.

\textit{Note on terminology: } Typically, “facilitator” refers to a physicist who either individually or with collaborators
engages directly with the audience in informal physics spaces. “Practitioner” refers to
a physicist who is involved in designing and managing an informal physics space; they
may or may not also act as facilitators in the space. For simplicity, we will use the
terms facilitator and practitioner interchangeably.

\subsection{Positionality}
As researchers, our affiliations and experiences in informal physics and physics education research influence the way we conducted this work. We include positionality statements to contextualize our findings because our backgrounds inevitably contain inherent biases, affordances, and limitations.

El-Adawy is a physics education researcher with expertise in STEM researchers’ professional development. She has conducted research on informal physics as part of her PhD. 
Lau is a physics education researcher with expertise in faculty professional development around teaching. She has been a facilitator of informal physics activities and currently manages a number of public engagement professional development programs. 
Sayre is a physics education researcher with expertise in persona generation in physics education spaces and STEM faculty professional development for teaching and research. She rarely engages in IPER directly, though she manages an award portfolio which includes informal physics. 
Fracchiolla is an IPER expert and facilitator, with more than 10 years experience designing, coordinating, and facilitating informal physics programs.

\section{Methodology: Personas}

Because we were trying to understand what are interests, needs, and challenges of APS members to do Public Engagement, we used a user-centered design methodology: personas. We use personas methodology because of its usefulness showcased in education research for instructional design and professional development. For example, the research team at PhysPort \cite{mckagan2020physport}, a professional development website for physics faculty, used personas  to improve the design and development of resources and activities for faculty by understanding their needs when making changes to their teaching \cite{madsen2020user}. Personas has also been used for undergraduate researchers to support the design of research programs with student-centered approaches based on their various motivations and experiences \cite{huynh2021building}. Additionally, personas has been applied to design instructional resources around learners' needs in the workplace \cite{maier2010using}. 

Personas are person-like constructs created from data of a group of potential users, which are synthesized into archetypes \cite{huynh2021building}. Each persona is created around a common goal for users that stems from the data and informs the design process. Data from multiple individuals are abstracted into one persona. Personas are typically created in sets which collectively represent the most important or frequent goals of users. Users can identify with multiple personas depending on their motivations, needs, and context. Personas allow us to create targeted professional development resources based on motivations, needs, and experiences of potential users because they highlight the diversity of potential users while centering their needs.  

By creating archetypes that are very human-like without representing the peculiarities of one person, several benefits emerge. 
First, researchers preserve the confidentiality of interview participants because the synthesized patterns are a combination of features from multiple interview participants \cite{madsen2020user}, a benefit for both researchers and research participants. Second, personas represent real users for which resources are meant to be created instead of the assumptions of designers who may envision a variety of resources that are not useful for the actual target population \cite{pruitt2003personas}, a benefit for users. Third, personas are person-like, and it is easier for designers to keep their needs in mind than it is to remember and relate to abstracted statistics about user segments, a benefit for designers.  Lastly, researchers focus on goals and needs of users across the entire design process of resources, which creates rich descriptions of a variety of experiences and needs of the target users, a benefit for all. These benefits to researchers, designers, users, and participants align with the goal of this project, which is to understand the needs of informal physics facilitators and develop user-centric resources to support them in order to lower barriers for informal physics education implementation and participation. 

\section{Framework: Self-Determination Theory}

We draw on Self-Determination Theory (SDT), a theory about motivation that centers around a learner’s agency when making choices to reach desired goals, \cite{ryan2000self} to examine, justify and interpret their motivation and needs. We use SDT \cite{ryan2000self} to inform  personas development as it allows us to hone in on individual motivation \cite{huynh2020professional}. Literature on informal physics education research supports that intrinsic motivation is a driving factor for engagement in informal physics for both facilitators and participants \cite{fracchiolla2016university}, so a theory which focuses on individual motivation is appropriate. SDT and personas methodology have been used together to identify research participants' various goals and motivations in previous physics education research work \cite{huynh2020professional}. Since our unit of interest in this study are individuals who have the opportunity to grow professionally in their interest and engagement with informal physics education, we deliberately choose SDT to investigate the motivation of physicists engaged in informal physics, centering around their agency in making choices to reach their goals.  SDT suggests that three psychological needs, competence, relatedness and autonomy, have to be satisfied to have the
most self-determined form of motivation \cite{stupnisky2018faculty}. We contextualized the definitions of the components to be applicable to the informal physics context as follows:
\begin{itemize}
    \item Autonomy: Desire to have sense of choice in their public engagement work;
    \item Relatedness: Desire to be connected and recognized with others in public engagement;
    \item Competence: Desire to experience mastery in public engagement work.
\end{itemize}

\section{Methods}
\subsection{Context: Recruitment of Research Participants}\label{informal physics context}
To identify individuals and networks engaged in informal physics to participate in our study, we used a snowball approach \cite{parker2019snowball}. We gathered an initial list of names and networks to tap into from researchers and practitioners in informal physics who engage with the American Physical Society (APS) public engagement efforts. Once we gathered a list of about 30 individuals, we sent out a screening questionnaire to ask if they were willing to participate in the research study and/or if they had suggestions for other individuals to seek out to broaden the network of practitioners and researchers we would interview. For those who responded positively to the screening survey, we then reached out to conduct one-on-one semi-structured interviews for our research study. 

Our data set contained 23 participants from various backgrounds and experiences in informal physics. Interviews were conducted with practitioners and/or researchers who are engaged in informal physics activities, events and programs. Interviewees covered a large span of career stages: graduate students, post-doctoral researchers, physics teachers, physics faculty, physicists at national or international labs, and science communicator professionals.  There is a large diversity on the type of activities and audiences they engaged in. Some of the activities included working with groups at universities, local schools, and a variety of public forums.   Though personas does not require data saturation \cite{guest2006many}, we note that many themes saturated across participant characteristics and informal physics activities.


\subsection{Data Collection}
The semi-structured interview protocol covered four main topics: (a) the interviewee's current role and experience with informal physics; (b) their conception of and motivation for informal physics work; (c) needs with informal physics work; and (d) professional identity. Our protocol included questions such as: 
\begin{itemize}
    \item Could you give us a broad overview of your current professional obligations and involvement in public engagement?
   \item What is your current informal physics education/research community?
   \item What are some challenges/barriers you are encountering with engaging in public engagement activities? 
   \item What would you need to overcome those challenges? What kind of support would be most helpful to you?
\end{itemize} 
Interviews were conducted by the first author over video conference (Zoom),  recorded, and transcribed (Zoom automatic transcription service) for analysis. The length of the interviews varied between 30 to 60 minutes depending on the availability of the interviewee and how much detail the interviewee gave in their answers.


\section{Personas Development}
We conducted a thematic analysis of the interview transcripts \cite{braun2006using}. The process consisted of iteratively reading the transcripts and paying particular attention to the participants' answers about goals, needs, and resources for engaging in informal physics. For each transcript, key ideas of participants were identified within emergent topics such as  \textit{interest in informal physics: recruiting underrepresented populations to physics}, \textit{resource used: discussions with practitioners}, \textit{challenge: isolated from community}, \textit{need: science communication: meeting your audience where they are at}.  
All transcripts were read and an initial list of topics was generated.   From this initial list of topics, we grouped and refined them into emergent themes.  We returned to the interviews with the emergent themes in mind to specifically seek evidence for them across the interview corpus and to develop and examine the breadth of human experience within each theme.  This process allowed us to refine the preliminary list of themes into broadly relevant and refined themes that emerged from our interview set.  Our thematic analysis generated emergent themes centered around participants' interests, resources used, challenges, and needs.  

For each interview, we additionally summarized biographical and demographic information about participants, including their career stage (physics graduate student, postdoc, faculty, informal physics professional, physicist at national laboratory, high school physics teacher). We connected this information to the themes present in each interview. 
Quotes illustrative of motivation to engage in informal physics were also included with this summary information for use in persona generation after thematic analysis.
 
Starting from the summarized information for each interview created from the thematic analysis, we used the components of Self-Determination Theory to differentiate the goals individuals may have. Using the self-determination theory (SDT) components (\textit{competence}, \textit{autonomy}, and \textit{relatedness}), we characterized participants and their motivations for engaging in informal physics to identify distinct patterns of motivations linked to resources our participants needed.  
For example, some participants are motivated by the opportunity to self-reflect on their journey and relate their growth to the public (motivated by \textit{relatedness}). They often expressed needs for resources around increasing their competence and mastery of their public engagement work, through skill development of varying degrees (resources to increase their \textit{competence}). 
Connecting motivations to needs via SDT allowed us to create a set of personas distinct in their goals and needs thanks to the framing offered by the theoretical components. 
As a design methodology, personas allow us to attend to these nuances of needs, which helps us to brainstorm ways to best support our goal of designing user-centered resources, improving the resources JNIPER can offer to better support the community of informal physics educators.  

The final step in generating personas is to enrich their motivations and resource needs with person-like features, such as names and avatars, to provide more context to the reader of activities and audience in our interview participants and to aid designers in using the personas to develop resources. Each persona's characteristics were drawn from the interview participants and their descriptions of their activities and engagement, blending features and activities across multiple participants.  Each persona therefore does not represent a single person, but a creative amalgam of participants.  This feature of personas, as a research methodology, allows us to preserve the confidentiality of the participants. 

We went through multiple iterations of thematic analysis and personas development. Discussions occurred among the research team on the personas created to refine their development until consensus among the project’s researchers was achieved. An early iteration of personas development is detailed in previous work \cite{el2022personas}. We include in this paper the details of the final iteration of personas development.

\section{Findings}

We present our set of three personas: Kyle, the self-reflective facilitator; Rory, the sparking interest and understanding facilitator; and Tracy, the representation matters facilitator. We summarize Kyle, Rory and Tracy's key needs and implications for the development of resources in Table \ref{personastable}.

\begin{table*}[t]\centering
\caption{Personas representing variation of physicists around needs in informal physics and potential implications for research team designing resources.} \label{personastable}

\begin{tabular}{p{0.2\linewidth} p{0.4\linewidth} p{0.4\linewidth}}

\parbox{\linewidth}{Persona} & \parbox{\linewidth}{Key needs} & \parbox{\linewidth}{Implication for designing resources}  \\
\hline
\parbox[t][][t]{\linewidth}{
\parbox[t][][c]{.28\linewidth}{\includegraphics[width=\linewidth]{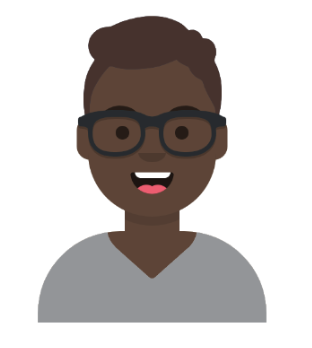}}
\parbox[t][][c]{.7\linewidth}{Kyle:\linebreak  \raggedright{the self-reflective facilitator}}}
&
\parbox[t][][t]{\linewidth}{
\begin{itemize}
\item A centralized resource hub to get started
\item Science communication training 
\item Skill development on how to organize to sustain engagement in informal physics
\end{itemize}
}
&
\parbox[t][][t]{\linewidth}{
\begin{itemize}
\item Designing  a searchable list of activities that are easy to implement
\item Designing training on skill development: storytelling with confidence and logistical programmatic factors
\end{itemize}
}
\\

\parbox[t][][b]{\linewidth}{
\parbox[c][][t]{.28\linewidth}{\includegraphics[width=\linewidth]{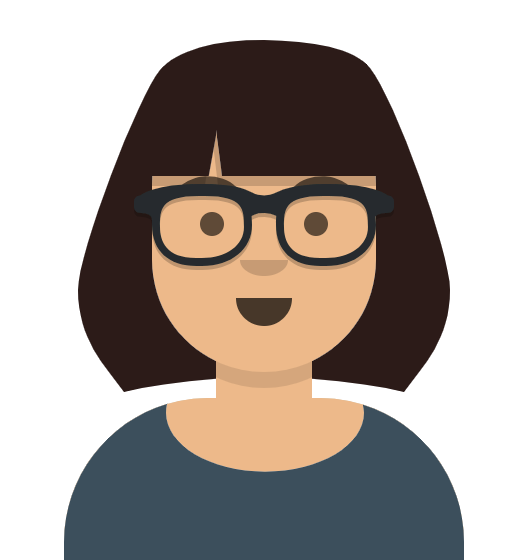}} 
\parbox[c][][t]{.7\linewidth}{Rory:\linebreak \raggedright{the sparking interest and understanding facilitator}}
}
&
\parbox[t][][t]{\linewidth}{
\begin{itemize}
\item Community building among physicists who do this work in isolation
\item Community building between physicists and science communication professionals
\end{itemize}
}
 & 
 \parbox[t][][t]{\linewidth}{
 \begin{itemize}
 \item Designing opportunities to share ideas and findings with other practitioners, professionals, researchers  at conferences
 \item Designing a network that allows practitioners to identify opportunities to partner with other practitioners or with researchers 
 \end{itemize}
}
\\

\parbox[t][][b]{\linewidth}{
\parbox[c][][t]{.28\linewidth}{\includegraphics[width=\linewidth]{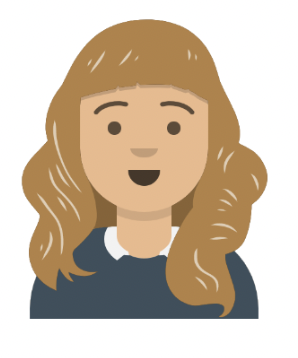}} 
\parbox[c][][t]{.7\linewidth}{Tracy:\linebreak \raggedright{the representation matters facilitator}}
}
&
 \parbox[t][][t]{\linewidth}{
 \begin{itemize}
\item Funding for informal physics
\item More buy-in from institutions
\item Investment in infrastructure to support informal physics
\end{itemize}
}
 & 
 \parbox[t][][t]{\linewidth}{
 \begin{itemize}
\item Designing spaces for discussions to occur to get the community to recognize and elevate the value of informal physics
\item Designing opportunities to share benefits of public engagement and advocate for funding
\end{itemize}
}
\end{tabular}
\end{table*}

\subsection{Kyle, the self-reflective facilitator}
Kyle, the self-reflective facilitator, engages in informal physics because they enjoy how energized they get when interacting with an audience to convey knowledge. Informal physics education activities are an opportunity to self-reflect on their experience in physics, especially their belonging and own understanding of content knowledge in physics, which allows them to relate their journey to their audience.
A representative quote of Kyle's goal is:
\begin{quote}
    \textit{I personally get a little bit of a high from doing it. I love to be in front of a crowd and talking about things that I know. I love answering people's questions.}
\end{quote}
Although engaging with the public energizes Kyle about their science and enables self-reflection, they find it challenging to figure out how to interact with different types of audiences. Kyle also faces organizational challenges. They are not sure how to best organize their engagement in informal physics to sustain their engagement for long periods of time while managing their many responsibilities. 

Kyle expressed needs around competence, particularly the desire to be better at informal physics. They would like to have access to centralized resources on how to get started when engaging with a specific type of audience or event in informal physics. They also would like to get training in science communication to best engage with different types of audiences and develop their skills in designing, managing and organizing activities and events with multiple stakeholders (volunteers, audience, institutions).

An example of Kyle would be a physics graduate student who is part of a student-led program that works with K-12 students during an after school program. They work to provide activities and illustrate physics concepts sometimes at the school or sometimes on university campus locations.  Kyle works with other graduate students in this program, which may have gotten started before they joined it, and Kyle's responsibilities in the program may vary from developing or implementing activities to coordinating with other facilitators.  

Based on the Kyle persona, we identified two approaches to support Kyle's resource and skill development needs (see Table \ref{personastable}). First, we need to design a searchable list of activities that are easy to implement when getting started with new content and new audiences. Second, we need to support the design of skill development which focus on storytelling while considering logistical programmatic factors. 

\subsection{Rory, the sparking interest and understanding facilitator}
Rory, the sparking interest and understanding facilitator, engages in informal physics because they enjoy conveying their excitement about science to others and seeing the ``light bulb" moments when participants understand a new physics concept. This motivation is driven by their desire to connect scientists and the public to form better relations and understanding of the scientific process. 
A representative quote of Rory's goal is:
\begin{quote}
    \textit{I love when students figure something out and they get super excited and start explaining it to all their friends. So the possibility that when I am doing one of these events that I could inspire someone to go to work in the sciences, to possibly work in physics areas that I am really passionate about. As a by-product, my work in outreach and engagement is also about getting the audience to appreciate science so the scientific process has become much more of what I try to teach.}
\end{quote}
As a result of sharing their excitement with their audience, Rory is not only hoping some participants may consider a STEM career path, but also appreciate the scientific process. Rory develops their skill as a facilitator through practice and trial and error. 

Rory's needs are centered around relatedness and connecting with the community, particularly being supported and engaged with a community of practitioners. They see two big gaps between groups of physicists around informal physics education, and they think it is important to bridge these gaps personally. The first gap is between physicists (e.g. faculty) who engage in informal physics and physicists who do not. The second gap is between full time informal physics professionals (e.g. full-time science communication professionals) and physicists who engage in informal physics part-time (e.g. physics faculty).  To bridge both of these gaps, Rory wants help to expand their engagement with the broader physics community towards a larger goal of elevating the perceived value of public engagement among physicists.

An example of Rory would be a science communicator professional who works closely with the public engagement units in national or international labs. Rory talks with the public during guided tours of the lab and plans demos for specific events for students at the lab. They also engage with the public on news outlets and radio shows about physics discovery, history or the latest newsworthy research developments.  Rory might have a personal podcast or other social media that centers science topics, or they might contribute to one as part of their job. 

Based on the Rory persona, we identified two community building approaches to support their needs. First, we need to create opportunities and spaces for various members of the informal physics community to interact with each other and support each other. Second, we need to design avenues for networking and building partnerships and collaborations among practitioners and researchers. 

\subsection{Tracy, the representation matters facilitator}
Tracy, the representation matters facilitator, engages in informal physics because of their identity connection with the audience. This motivation is driven by the value they see in inspiring diverse people to pursue STEM careers paths.   
A representative quote of Tracy's goal is: 
\begin{quote}
    \textit{I'm trying to get more girls, women, and people of color into physics.}
\end{quote}
Tracy discusses their informal physics efforts with other practitioners but is frustrated by the pushback they receive from the physics community, which does not always see it as an integral part of a physicist's job. To support their work in this space, they need resources to foster their autonomy:
\begin{itemize}
    \item Funding to allow them to recruit and retain more individuals in informal physics programs as both participants and facilitators; and to expand assessment of programs and informal physics events;
    \item  More buy-in from institutions on the value of their informal physics work, which would foster their sense of agency in what they can do in this space;
    \item Logistical and managerial  support for their public engagement activities. They need more infrastructure to be built in order to foster their sense of autonomy. This will allow them to dedicate their time and effort to the content and design of the engagement activities. 
\end{itemize}
Tracy's motivation and needs are congruent with findings from the literature, which has shown the critical role that recognition and relational resources play in linking programmatic efforts to support representation of students from underrepresented groups and physics identity development \cite{hyater2018critical}.

An example of Tracy would be a physics faculty who engages with the general public during public talks about their science. They might work with K-12 schools to provide information and illustrate how some physics concepts work through a series of demonstrations. 
Alternately, Tracy might participate in events with youth organizations around science topics (e.g. STEM badges for Girl Scouts).  

Based on the Tracy persona, we identified two resources to support them to advocate for their work in this space. First, we need to design space for discussions to occur to get the physics professional community to recognize and elevate the value of informal physics activities. Second, we need to design opportunities to share the benefits of public engagement and advocate for funding for informal physics programs. 

\section{Discussion}
The focus of this study was to better understand facilitators' motivations and professional needs in informal physics. From our dataset,  we generated three personas: the physicist who engages in informal physics for self-reflection, the physicist who wants to spark interest and understanding of physics, and the physicist who wants to provide diverse role models to younger students and inspire them to pursue a STEM career.  Using personas highlighted features of physicists' needs we may not have captured otherwise, or may not have centered in developing materials. For example, we might have thought that materials should be aimed at different career stages for physicists: materials for graduate students, for faculty, for full-time science communicators, etc.  However, in constructing these personas, we noticed that career stage and motivation are not in a one-to-one correspondence. There were multiple career stages represented in each persona, and the needs of facilitators mapped better to their motivations than their career stages.  Foregrounding facilitators' motivations allows us to design materials that are better at meeting the needs of diverse informal physics educators.  Personas, as a research methodology, coupled to SDT, as a theoretical framework, allowed us to take a user-centered approach.  

By developing this set of personas, we expand on the informal physics community's understanding of the needs of practitioners in this space. 
The development of these three personas informs the design of resources listed in the third column of Table \ref{personastable}, created for JNIPER. 
This list of possible resources informed the first set of initiatives APS JNIPER program launched in Fall 2022, which includes monthly coffee hours and a JNIPER slack channel where members share resources. The coffee hours address Rory's need around community building, specifically the need of connecting several types of professionals in the informal physics space. The coffee hours also address Tracy's need to have discussion spaces to advocate for the needs of informal physics facilitators' autonomy.  Furthermore, the active online community is a first step in addressing Kyle's need of having a resource hub on how to get started, share best practices and materials. As expected with personas methodology, a few activities can serve multiple user-types, even if the reason why the activity is helpful differs between each user. Hence, bringing this methodological approach to  professional development in informal education enriches the development of user-centric resources to support informal physics facilitators. 

Additionally, our findings can support other facilitators in informal physics programs because our personas could help them better articulate their needs and identify mechanisms to address them. While this research study provided a baseline for programmatic design for APS Public Engagement, the needs and challenges are representatives of individuals across the informal physics spectrum. Therefore, departments, organizations and institutions could draw upon the personas developed to consider the ways to better support physicists in their respective environment. 

Moreover, our findings underline an important feature of public engagement, which is that facilitation of public engagement is driven by intrinsic motivation. This theme that previous research had articulated as a significant contributor of participation in informal physics programs \cite{fracchiolla2016university}, emerged in our personas as well. It is important to highlight this self-driven underlying motivation. Participation in any capacity in informal physics is often voluntary and internally motivated, and prior work has centered the freedom to explore and engage with physics.  Our personas showcase participants' intrinsic fulfilment in connecting with one's journey and with others when engaging in informal physics.  These important motivational aspects to engaging in informal physics emerge directly from our analysis and persona development. 

Given JNIPER's aim to support informal physics endeavors by grounding the network's resources in research, equity, and culturally competent practices, future work should complement our work on facilitators by examining the motivations and needs of participants. We designed our study to focus on facilitators' experiences and resource needs due the minimal research on informal physics facilitators. However, at its core, public engagement is a two-way interaction that requires professional development resources that support both parties. Consequently, in the future we should consider how facilitators' professional development resources interact with the motivations and resource needs of participants in informal physics. This holistic approach could allow us to form a network that is culturally responsive and inclusive of all stakeholders in the informal physics space.

\section{Conclusion}

Physicists engage with the public to varying degrees at different stages of their careers, but their public engagement covers many activities, events, and audiences, making their motivations and professional development needs not well understood. As part of ongoing efforts to build and support community in the informal physics space, this paper discussed the findings from our interviews with physics practitioners and researchers with a range of different experiences in informal education. These findings support our larger goal to design useful and targeted resources for stakeholders. We discussed our personas development process where we determined existing interest and professional development needs of practitioners and researchers in this space. The development of the three personas brings user-centered design to informal physics professional development research. It also broadens our understanding of motivations and needs of physicists engaged in science outside the formal classroom.

\begin{acknowledgments}
This work is supported by the American Physical Society. 
\end{acknowledgments}



\bibliography{apssamp}

\end{document}